\documentclass[aip,reprint,english]{revtex4-1}
\pdfoutput=1
\usepackage{amsmath}
\usepackage{color}
\usepackage{graphicx}
\usepackage{epstopdf}

\usepackage{babel}

\begin{document}

\title{Structural properties of Co$_{2}$TiSi films on GaAs(001)}

\author{B.~Jenichen}
\email{bernd.jenichen@pdi-berlin.de}
\author{J.~Herfort}
\author{M.~Hanke}
\author{U.~Jahn}
\author{X.~Kong}
\author{M.T.~Dau}
\altaffiliation[Now at ]{INAC-SPINTEC, CEA/CNRS and Universit$\acute{e}$ Grenoble Alpes, F-38000 Grenoble, France}
\author{A.~Trampert}
\address{Paul-Drude-Institut f\"ur Festk\"orperelektronik,
Hausvogteiplatz 5--7, D-10117 Berlin, Germany}
\author{H.~Kirmse}
\affiliation{Humboldt Universit\"at  Berlin, Institut f\"ur Physik,
Newtonstrasse 15, D-12489 Berlin,  Germany}
\author{S.~C.~Erwin}
\address{Center for Computational Materials Science,
Naval Research Laboratory,
Washington, DC 20375, USA}

\date{\today}

\begin{abstract}
Co$_{2}$TiSi films were grown by molecular beam epitaxy on GaAs(001) and analyzed using reflection high-energy electron diffraction, and electron microscopy. In addition, X-ray diffraction was combined with lattice parameter calculations by density functional theory comparing the \textit{L$2_1$} and \textit{B}2 structures and considering the influence of non--stoichiometry. Columnar growth is found and attributed to inhomogeneous epitaxial strain from non-random alloying. In films with thicknesses up to 13~nm these columns may be the origin of perpendicular magnetization with the easy axis perpendicular to the sample surface. We found \textit{L$2_1$} and \textit{B}2 ordered regions, however the [Co]/[Ti]--ratio is changing in dependence of the position in the film. The resulting columnar structure is leading to anisotropic \textit{B}2--ordering with the best order parallel to the axes of the columns.
\end{abstract}



\maketitle

\section{Introduction}
Ferromagnetic (FM) Heusler alloys are compatible with semiconductor (SC) technology and may be useful as electrodes for spin injection into the semiconductor.\cite{Felser2005,ramsteiner08,Graf2011,Bai2012, Bruski2013,Manzke2013,Felser2015}
Among them Co$_{2}$TiSi is well suited for spincalorics, i.e. the generation of spin currents in ferromagnets by the thermoelectric effect.\cite{Chen2006b} Co$_{2}$TiSi is expected to be half--metallic and thus promising a high degree of spin-polarization. The properties of bulk Co$_{2}$TiSi are described in detail in Ref.~[\onlinecite{Barth2010}]. Accordingly the lattice parameter calculated by density functional theory (DFT) was obtained for the completely \textit{L$2_1$}--ordered lattice to be a$_{DFT}$~=~5.758~\AA~  and the measured lattice parameter is a$_{exp}$~=~5.849~\AA~, i.e. the theoretical lattice misfit of a$_{DFT}$ with respect to the GaAs substrate a$_{GaAs}$~=~5.653~\AA~ is near 2~\%.\cite{Barth2010}  The Seebeck coefficient of bulk Co$_{2}$TiSi is -30~$\mu$V/K at room temperature. The Curie temperature is approximately 300~K.\cite{Barth2010} Similar values of the Seebeck coefficient and the Curie temperature are found for Co$_{2}$TiSi thin epitaxial films grown by molecular beam epitaxy (MBE).\cite{Dau2015}

\begin{figure}[!t]
\includegraphics[width=8.0cm]{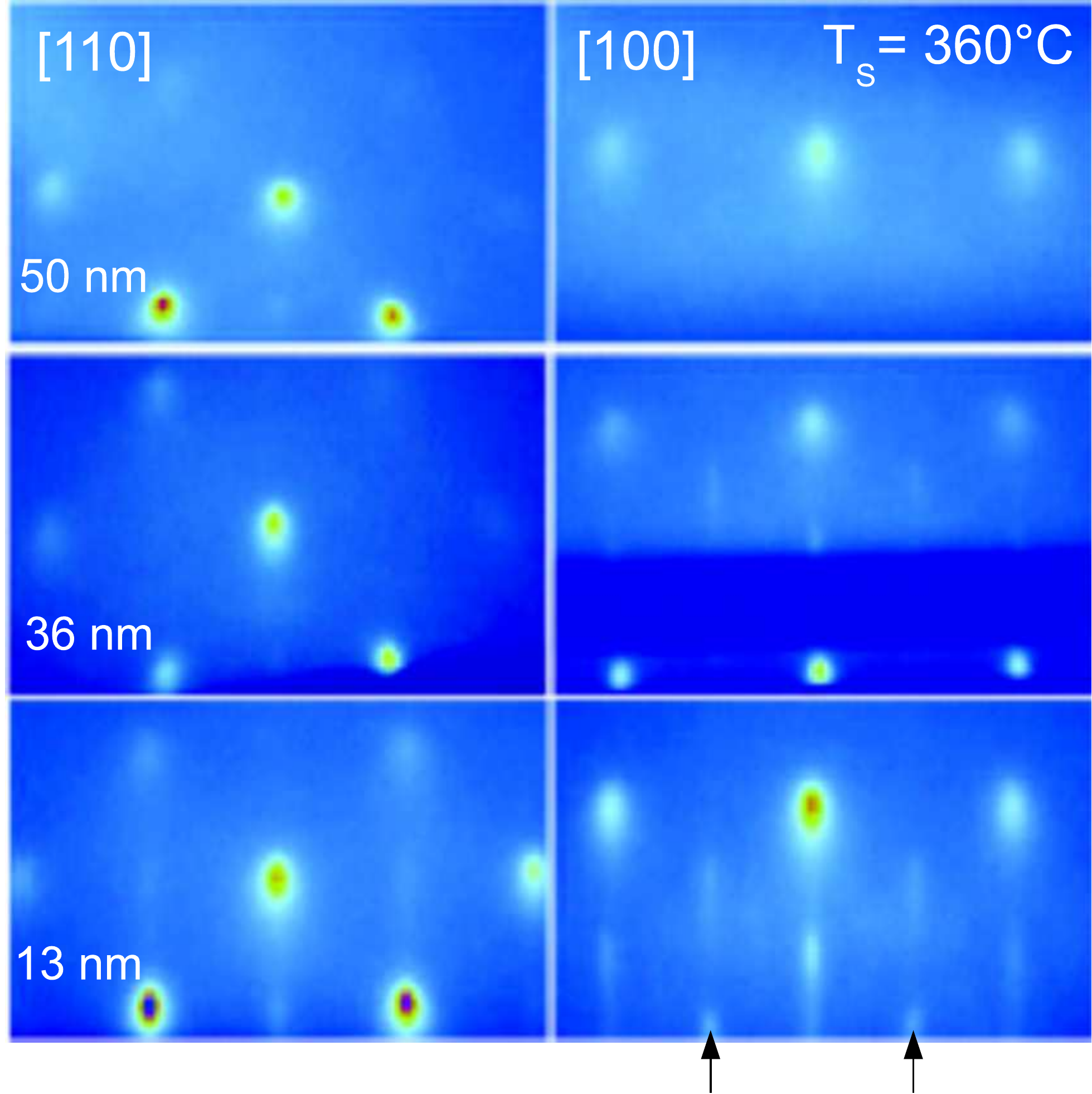}
\caption{(color online)  RHEED patterns along the [110] and the [100] azimuths for different thicknesses of the Co$_{2}$TiSi film. The superstructure maxima are marked by arrows.
}
\label{fig:rheed}
\end{figure}

\begin{figure}[!t]
\includegraphics[width=8.5cm]{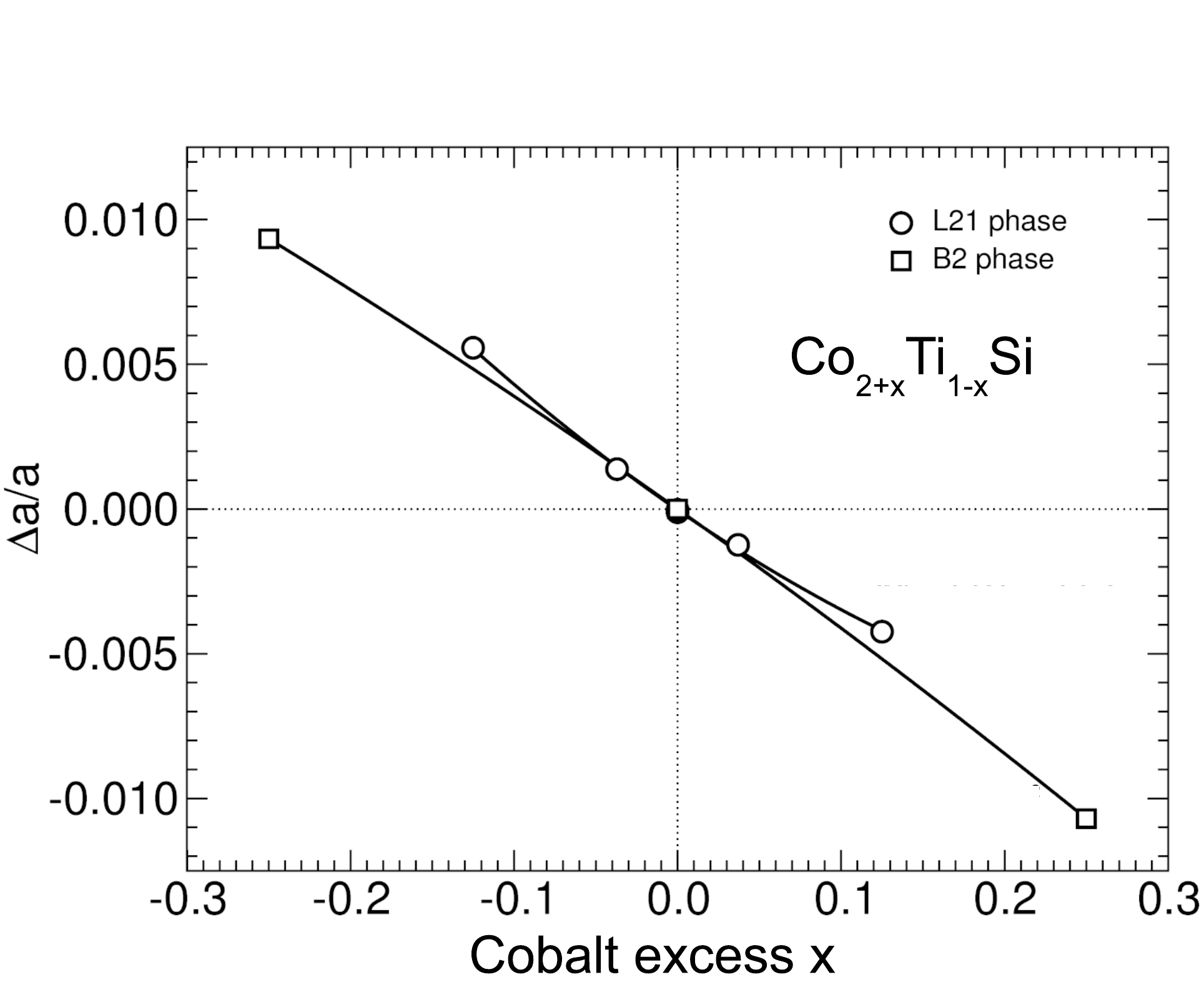}
\caption{Influence of the deviation x of stoichiometry of a Co$_{2}$TiSi film on the lattice parameter change calculated by DFT for the two different types of ordering \textit{L$2_1$} and  \textit{B}2. The approximations with deviations from linearity are given in corresponding formulas
(for the \textit{$L2_1$} ordering we obtain: $\Delta$a/a~=~-0.038~x~+~0.045~x$^{2}$,
for \textit{B}2 ordering we obtain:~$\Delta$a/a~=~-0.040~x~-~0.011~x$^{2}$).
The origin corresponds to stoichiometric Co$_{2}$TiSi.
}
\label{fig:dft_result}
\end{figure}

\begin{figure}[!t]
\includegraphics[width=7.5cm]{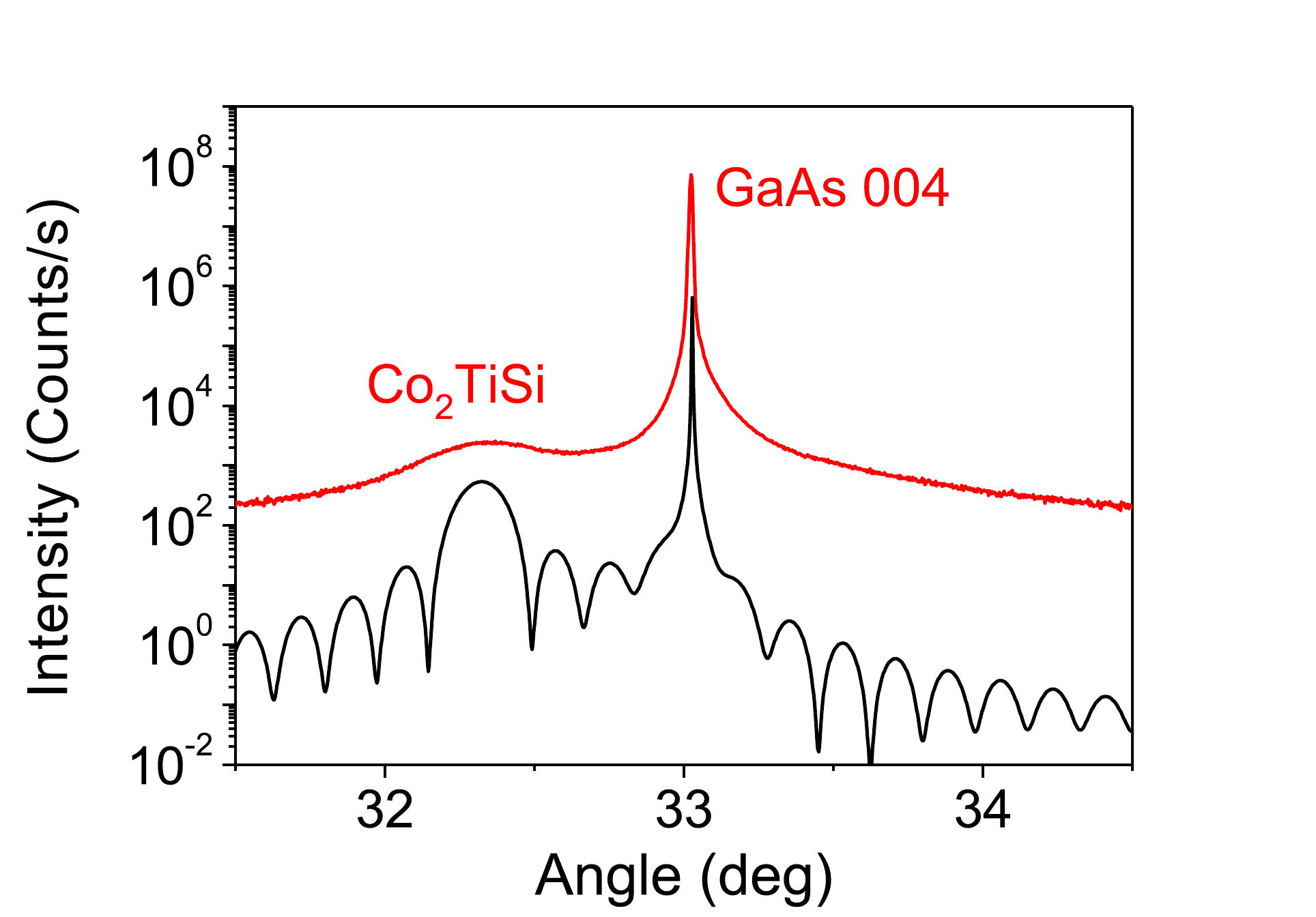}

\caption{(color online) Out--of--plane X-ray diffraction scan using the symmetrical (004) reflection of sample~1, comparison between experiment (above) and simulation (below, shifted for clarity). In the simulation we assumed a lattice parameter difference of $\Delta${a}/a~=~0.0195 between the film and the substrate in order to obtain a coincidence between the corresponding peak positions.
}
\label{fig:xrd004}
\end{figure}

\begin{figure}[!t]
\includegraphics[width=10.5cm]{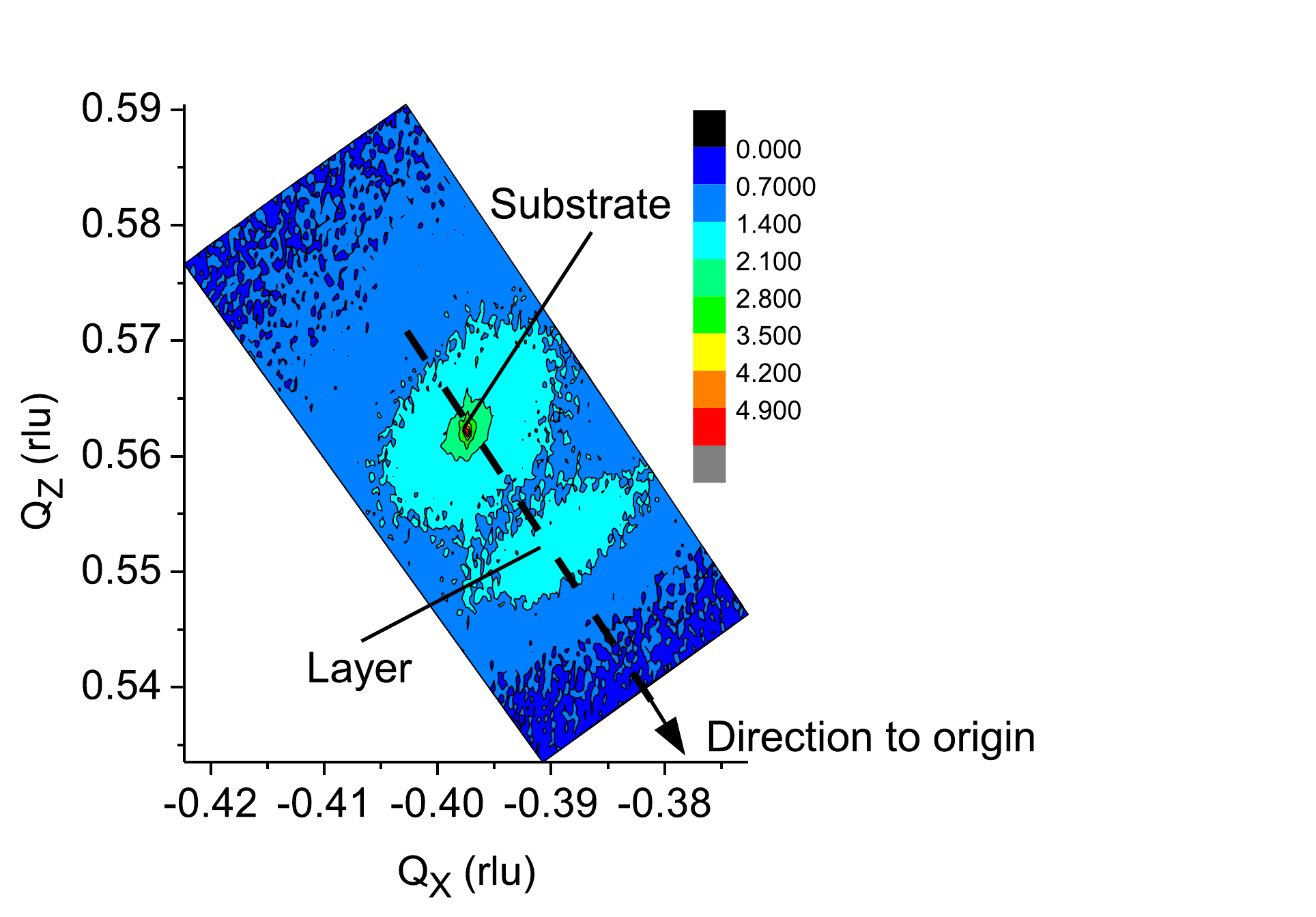}
\caption{(color online) Out--of--plane X-ray reciprocal space map using the asymmetrical (224) reflection of sample~1. The diffracted intensity is plotted on a logarithmic scale. The arrow is pointing to the origin of the reciprocal space.
}
\label{fig:rsm224}
\end{figure}


\begin{figure}[!t]
\includegraphics[width=7.5cm]{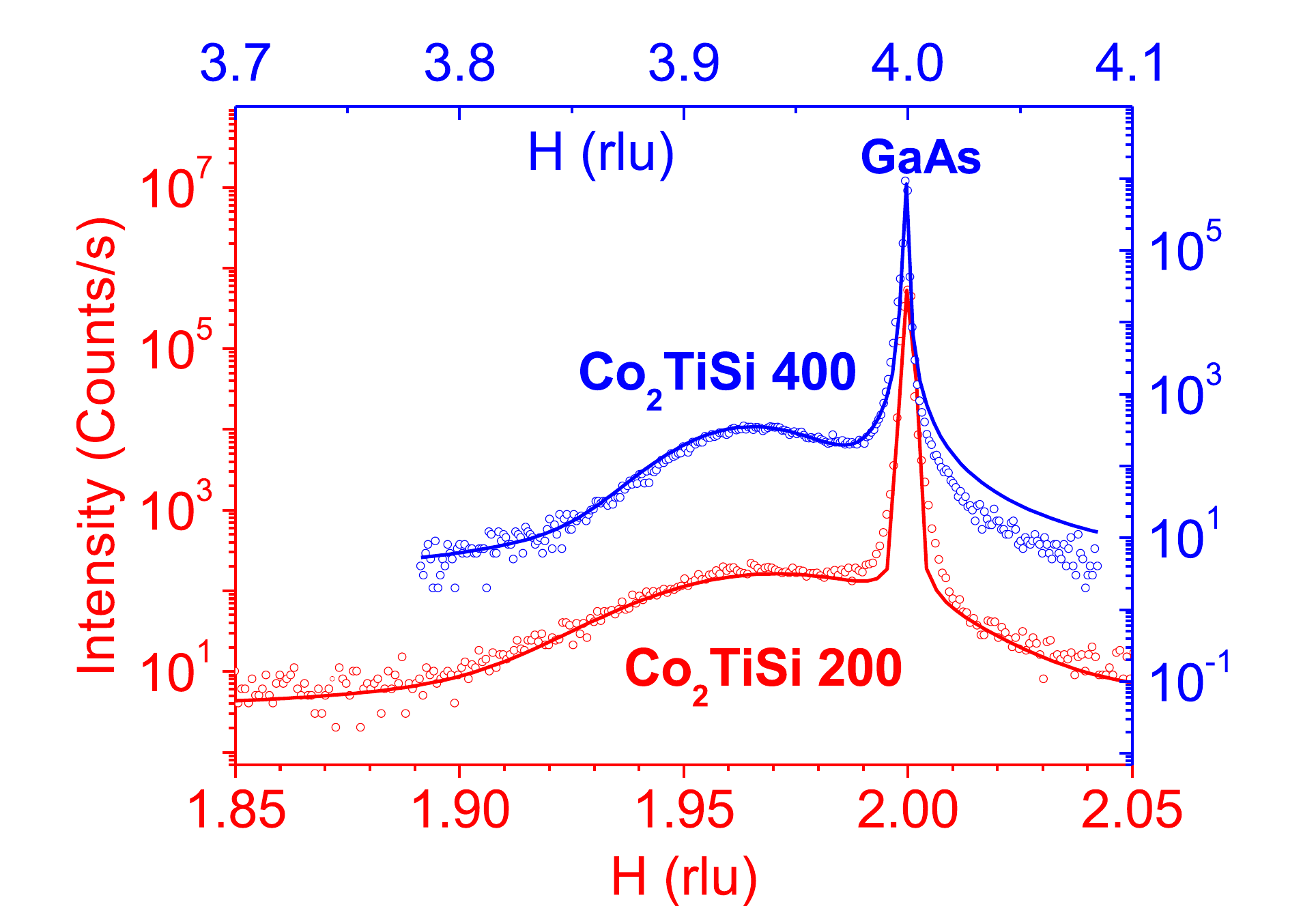}

\caption{(color online) In--plane H-scans of sample~1 using (400) and (200) reflections in grazing incidence diffraction (circles: experimental values, lines: fits by Voigt--functions).
}
\label{fig:xrd_hscans}
\end{figure}

\begin{figure*}[!t]
\includegraphics[width=15cm]{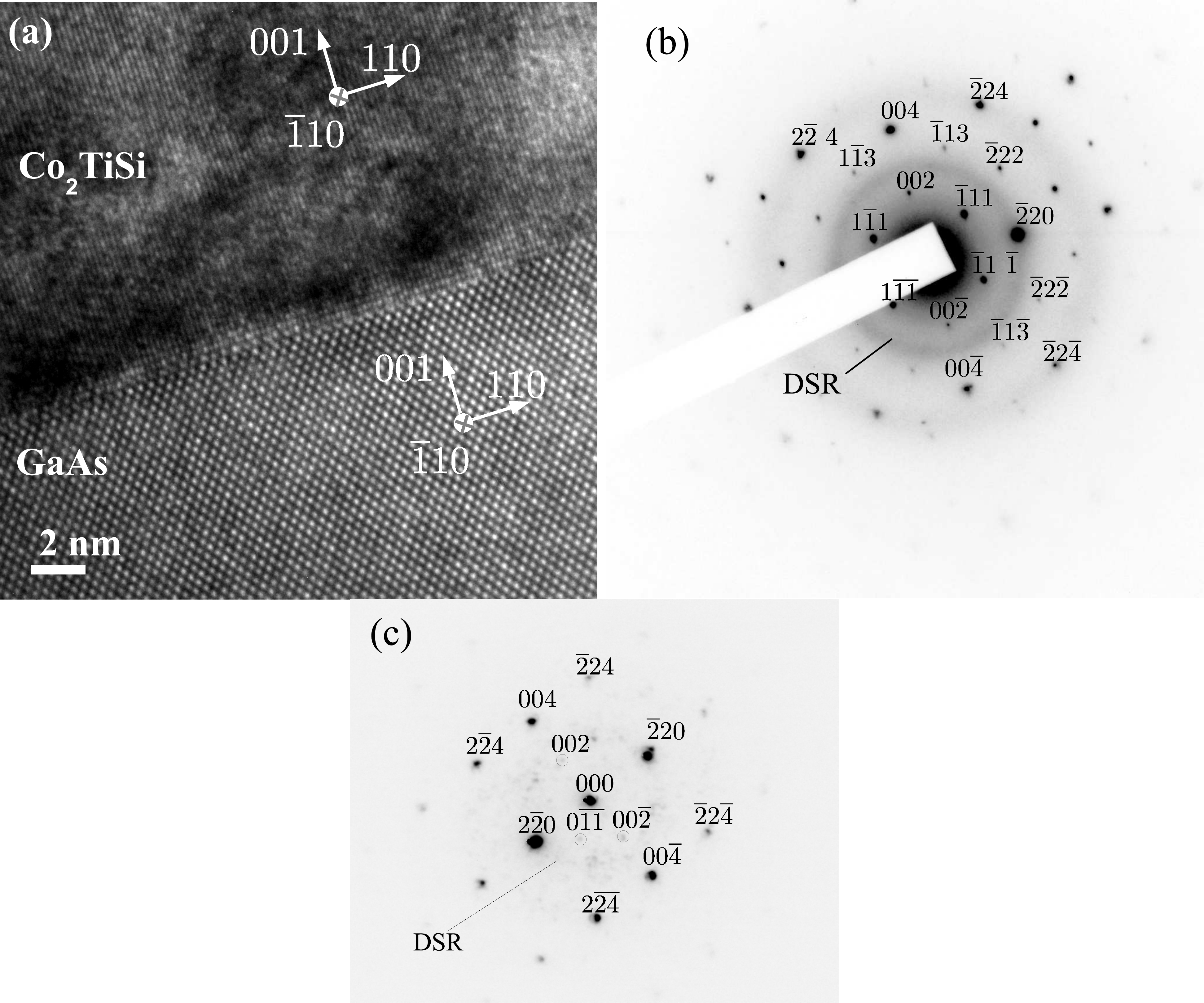}

\caption{(a) HR TEM micrograph of an almost perfect region near the Co$_{2}$TiSi/GaAs interface. (sample~1) The crystallographic orientations of film and substrate coincide. (b) Selected area diffraction pattern of the Co$_{2}$TiSi film. (sample~1) Debye-Scherrer rings (DSR marked by thin lines)  are visible through the fundamental maxima ($\overline{2}$20) and ($\overline{2}$24). These rings are probably caused by amorphous areas of the sample. (c) The corresponding nanobeam diffraction (NBD) pattern. Here the DSR (marked by thin line) appears spotty and not all superlattice reflections are clearly visible due to low intensity. Some of the faint superlattice reflections are marked by circles.
}
\label{fig:sad1}
\end{figure*}

\begin{figure*}[!t]
\includegraphics[width=15cm]{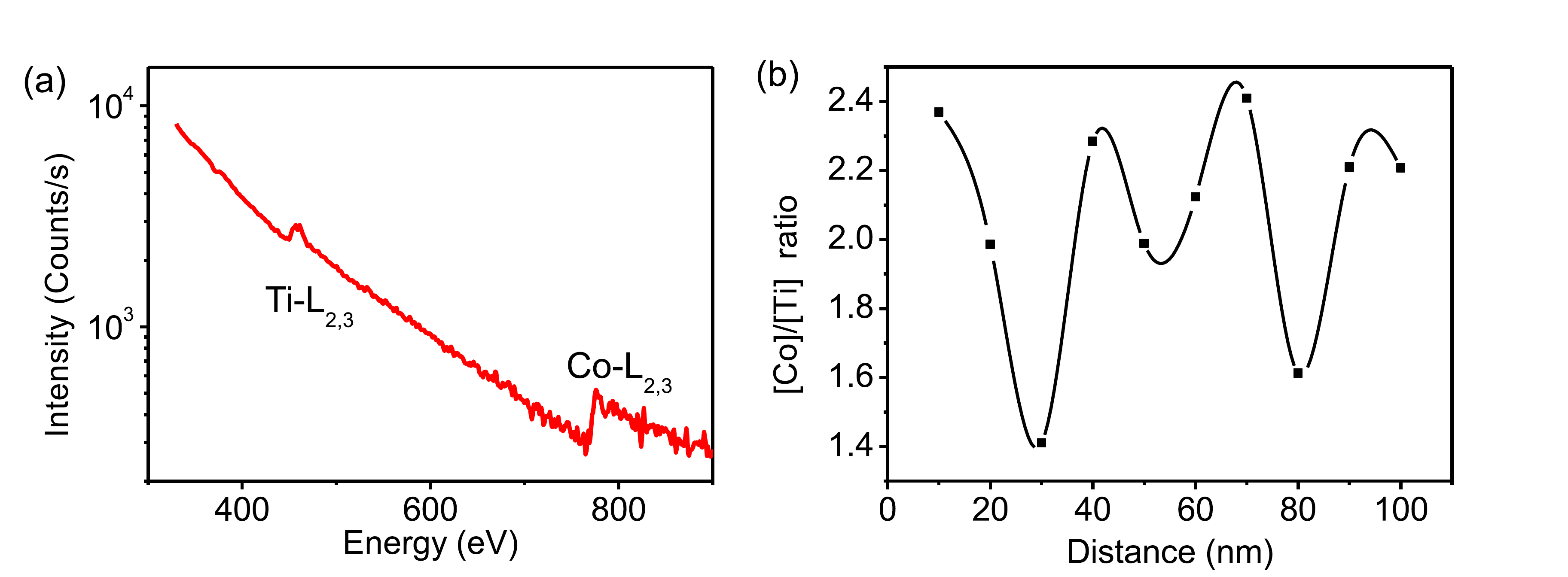}
\caption{(color online)(a) EELS spectrum of sample~2 and (b) plot of lateral distribution of the [Co]/[Ti]-ratio. The Ti-L$_{2,3}$ and the Co-L$_{2,3}$ edges are visible in (a).
}
\label{fig:eels}
\end{figure*}

\begin{figure*}[!t]
\includegraphics[width=15cm]{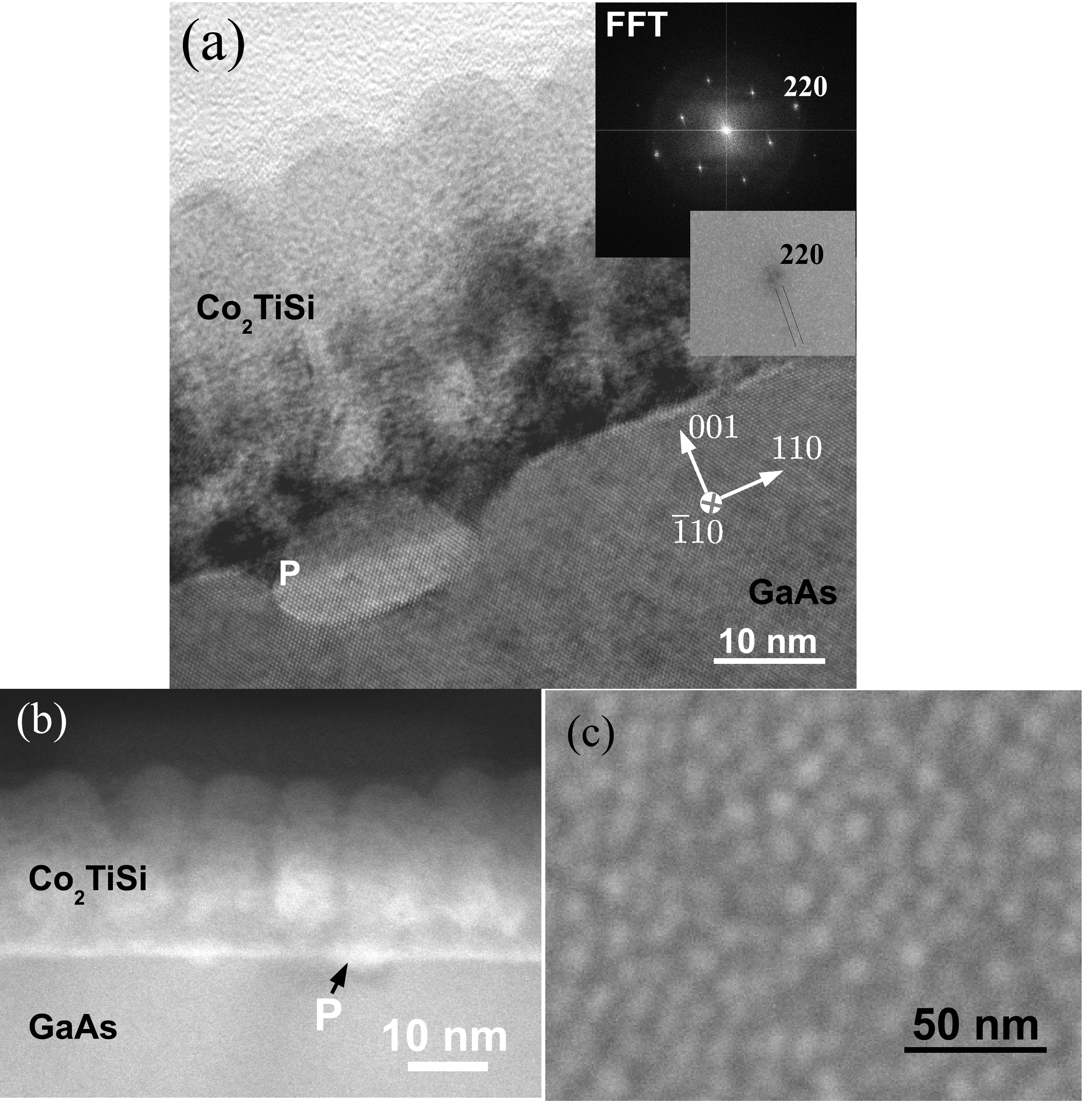}
\caption{ (a) Multi-beam HR TEM micrograph of a Co$_{2}$TiSi film (cross--section) of sample~1. The inset shows the fast Fourier transform (FFT) of the image and the inverted image is a magnification of the (220) maximum. Precipitates are marked by P. (b) STEM HAADF electron micrograph of a Co$_{2}$TiSi film (cross--section). (c) SEM secondary electron micrograph of the surface (top view).
}
\label{fig:hrtem1}
\end{figure*}

\begin{table}
\caption{Growth temperature $T_S$, film thickness (nominal thickness), and percentage of average long-range order S$_{B2}$  of Co$_{2}$TiSi films grown on a GaAs(001) substrate at two different $T_S$} \vspace{12pt} \label{tab:data}
\begin{tabular}{ccccc}
~~sample ~~&~~~$T_S$~~~& ~~~thickness~~~ & ~~S$_{B2}$~~~& ~~S$_{B2}$    \\
~~number~~&    ($^{\circ }$C)& (nm) &  out-of-plane~~ & in-plane~~\\
\hline
 1 &360 & 35$\pm$2 (35) &  35\% &  2\%  \\
 2 & 300 & 32$\pm$2 (35) & 20\% &  2\%  \\
\end{tabular}
\label{tab:tab1}
\end{table}

In addition, on GaAs(001) substrates a perpendicular magnetization of the Co$_{2}$TiSi film was found.\cite{DauAIP2015} Magnetization curves of similar Co$_{2}$TiSi films with different thicknesses (between 9~nm and 36~nm) have been measured at low temperature (T~=~10~K) along the directions perpendicular and parallel to the sample surface.  For layer thicknesses below 13.5~nm the easy axis of magnetization is directed perpendicular to the sample surface, whereas for the thicker films the easy axis is parallel to the surface.\cite{DauAIP2015} As this perpendicular magnetic anisotropy was shown to persist up to room temperature, it could not solely  be ascribed to the presence of interface clusters.
Here a Volmer-Weber growth mode of the Co$_{2}$TiSi film on GaAs would be possible, driven by epitaxial strain and poor wetting of the semiconductor by the metal.\cite{Bauer1958,koch94,kag09}
In the present work we investigate in detail the film structure and the homogeneity of the films. We give a possible explanation for the perpendicular magnetization achieved for lower film thicknesses. Non-random alloying was found in magnetic semiconductor thin films highly doped
with transition metals.\cite{Dietl2015} We suggest that similar phenomena may occur in thin films of magnetic Heusler alloys as well.

\section{Experiment and Calculation}

GaAs(001) buffer layers exhibiting the As-terminated c(4$\times$4) reconstructed surface were prepared by MBE.
Co$_{2}$TiSi films on GaAs(001) were grown in an As-free chamber for metal growth, which is directly connected to the III-V growth chamber via UHV. The epitaxial growth was performed at different substrate temperatures T$_{S}$ from 100~$^\circ$C up to 380~$^\circ$C as described in detail in Ref. \onlinecite{Dau2015}. With the help of a careful combination of X-ray diffraction (XRD), energy-dispersive X-ray spectroscopy (EDX) and electrical measurements we proofed the stoichiometric composition of the films as described previously.\cite{hashimoto05,Dau2015}.

The MBE growth was analyzed \textit{in-situ} using reflection high--energy electron diffraction (RHEED).\cite{DauAIP2015} Snapshots of RHEED patterns are given in Fig.~\ref{fig:rheed}. They were taken in the two different azimuths [110] and [100] for different film thicknesses 13~nm, 36~nm and 50~nm of the growing Co$_{2}$TiSi film. Cube on cube growth is observed with the epitaxial relationship ([110]Co$_{2}$TiSi $\|$ [110]GaAs). A tendency towards streaky RHEED patterns is found for film thicknesses below 15.0 nm.  Along the [100] azimuth a (2$\times$1) Co$_{2}$TiSi surface reconstruction is occuring during deposition up to a film thickness of 36~nm.  Distinct RHEED maxima without streaks appear with increasing thickness and a diffuse background is found, i.e. a roughening of the surface is occuring during the growth.    Two of the samples of the series with the same nominal thickness of 36~nm were chosen for further characterization, sample~1 grown at a substrate temperature  T$_{S}$~=~360~$^\circ$C and sample~2 at T$_{S}$~=~300~$^\circ$C (cf. Table~\ref{tab:tab1}).
The resulting structures were characterized by XRD, transmission electron microscopy (TEM), and scanning electron microscopy (SEM).

High-resolution XRD and X-ray reflectivity measurements were performed using a Panalytical X-Pert PRO MRD\texttrademark\ system with a Ge(220) hybrid monochromator  (CuK$\alpha_1$ radiation with a wavelength of $\lambda=1.54056$~\AA).
XRD patterns were calculated in dynamical approximation \cite{Stepanov1997}.
Some of the X-ray measurements were performed in grazing incidence geometry at the PHARAO U-125/2 KMC beamline of the storage ring BESSY II in Berlin. The photon energy was 10~keV, with an energy resolution $\Delta{E}/E\sim$10$^{-4}$.
XRD reciprocal space mapping (RSM) was performed for both samples in order to measure the degree of relaxation $\xi$ of the Co$_{2}$TiSi film on the GaAs buffer layer and substrate.\cite{Heinke1994}

The TEM specimens were prepared in the usual way by mechanical lapping and polishing, followed by argon ion milling.
High-resolution (HR) TEM images and selected area diffraction (SAD) patterns were acquired with a JEOL 3010 microscope operating at 300~kV. The cross-section TEM methods provide high lateral and depth resolutions on the nanometer scale. Electron energy loss spectroscopy (EELS) was performed in the TEM. The scanning TEM (STEM)  JEM--2200 FS microscope operating at 200~kV was used for EDX,  high-angle annular dark field (HAADF) imaging and nano beam diffraction (NBD).
In addition the samples were investigated by SEM, especially using secondary electron (SE) images and also electron backscattered diffraction (EBSD) orientation maps.


We calculated the lattice parameter of Co$_{2}$TiSi by density functional theory (DFT).\cite{Jones2015}  In this way the equilibrium state of the Co$_{2}$TiSi as a function of composition can, in principle, be obtained.
DFT in the generalized gradient approximation\cite{Perdew1996} was applied in order to determine the lattice parameter of the Co$_{2}$TiSi in dependence on the composition for the two different types of ordering \textit{B}2 and \textit{$L2_1$} using the Vienna Ab Initio Simulation Package (VASP).\cite{Kresse1996A,Kresse1996B}

\section{Results and Discussion}

 As a result of our DFT calculations we obtain the lattice parameter for stoichiometric Co$_{2}$TiSi with the \textit{$L2_1$} ordered lattice a$_{DFT}$(\textit{$L2_1$})~=~5.76~{\AA} and for the \textit{B}2 ordered lattice a$_{DFT}$(\textit{B}2)~=~5.80~{\AA}, i.e. the lattice parameters for the different types of ordering are slightly different. The reason for the dependence of the lattice parameter on the ordering is, that the bond--lengths of Ti--Si pairs and the Ti--Ti or the Si--Si pairs are not identical.
Figure~\ref{fig:dft_result} shows the influence of the stoichiometry of a Co$_{2+x}$Ti$_{1-x}$Si film on the lattice parameter change $\Delta${a}/a, where \textit{x} is the excess of Co in comparison to the stoichiometric composition ($x~=~0$). The dependencies are almost linear, although with slightly different slopes for the \textit{$L2_1$}
and the \textit{B}2 ordering. We now can compare these calculated values with the experimental lattice parameters measured by XRD. We need to measure the lattice parameters perpendicular and parallel to the interface (IF).

Figure~\ref{fig:xrd004}  shows an out-of-plane XRD scan of sample~1 using the symmetrical (004) reflection with a comparison between experiment (above) and simulation for an ideal Co$_{2}$TiSi film (below). Assuming a fully relaxed lattice of the Co$_{2}$TiSi we obtained a lattice parameter of the film of a$_{exp}$~=~(5.764$\pm$0.005)~{\AA}.  This value is nearly equal to the calculated lattice parameter of the \textit{L$2_1$} ordered lattice and lower than the lattice parameter of the \textit{B}2 ordered and fully stoichiometric films. This result is influenced by the average stoichiometry and the degree of ordering of the present epitaxial film. The broadening of the experimental film peak is due to misfit dislocations and other defects near the Co$_{2}$TiSi/GaAs IF, the IF-- and surface--roughness, as well as the inhomogeneity of the film. For the determination of the lattice parameter we needed to check the validity of our assumption, i.e. the degree of relaxation of the film with respect to the substrate lattice.
Figure~\ref{fig:rsm224} depicts an X-ray RSM of sample~1 near the fundamental maximum (224). The maximum of the layer peak lies on a line connecting the substrate peak and the origin of the reciprocal lattice (see arrow). This means that the film is fully relaxed. Similar results were obtained for both samples using different asymmetric reflections. In addition we have also measured in--plane reflections in order to check the anisotropy of the film properties.

In Figure~\ref{fig:xrd_hscans} in-plane reciprocal lattice H-scans of a Co$_{2}$TiSi film on GaAs(001) (sample~1) are given. The experimental peaks are fitted by Voigt-functions.\cite{Amstrong1967} The lattice mismatch between the Co$_{2}$TiSi film and the GaAs buffer layer is the same as in the corresponding out-of-plane scan given in Fig.~\ref{fig:xrd004} (a), i.e. the film is fully relaxed with respect to the GaAs buffer layer. The intensities of the reflections \{200\} and \{400\} are compared for determination of the \textit{B}2 order of the film.\cite{Jenichen2012}  Due to the low intensity of the superlattice reflections corresponding to the \textit{L$2_1$} ordering unfortunately the maxima are hardly detected by grazing incidence XRD and cannot be used for quantitative analysis of \textit{L$2_1$} ordering.

Table~\ref{tab:tab1} reports the results for the long-range order of the films. It shows the growth temperatures $T_S$, film thicknesses (nominal thicknesses), and percentages of average long-range order S$_{B2}$  of Co$_{2}$TiSi films grown on a GaAs(001) substrate at two different $T_S$. The degree of ordering S$_{B2}$~\cite{Jenichen2012} is increasing with growth temperature $T_S$.  The ordering along the surface normal is by an order of magnitude higher than the in-plane ordering, probably a consequence of columnar film growth. The measured intensities were weighted by the observed increase of the quasi--forbidden GaAs(002) reflection due to strain near the GaAs/Co$_{2}$TiSi IF. We postulated that the corresponding Co$_{2}$TiSi(002) reflection was increased by the same factor as GaAs(002). Below we investigate the structure of the films and the Co$_{2}$TiSi/GaAs IFs also on a smaller length scale by HR TEM.

Figure~\ref{fig:sad1}~(a) reveals a TEM micrograph of a region near the Co$_{2}$TiSi/GaAs IF of sample~1 and Fig.~\ref{fig:sad1}~(b)  a correponding SAD pattern of the Co$_{2}$TiSi film. The micrograph demonstrates a high quality of the IF although some inhomogeneity is detected inside the film. In the SAD pattern Debye-Scherrer rings (DSR) are visible intersecting the Co$_{2}$TiSi fundamental maxima ($\overline{2}$20) and ($\overline{2}$24).\cite{takamura09} As expected the fundamental peaks of the Co$_{2}$TiSi film are stronger than the superlattice maxima.\cite{Jenichen2012} The DSR may be partly connected to the glue used during sample preparation. Figure~\ref{fig:sad1}(c) shows the corresponding nanobeam diffraction (NBD) pattern obtained with a beam diameter of 0.5~nm. Here the DSR appears spotty and not all superlattice reflections are clearly visible due to low intensity. The SAD and NBD give evidence for \textit{L$2_1$} and \textit{B}2 ordering of the lattice, because diffraction maxima like \{111\} and \{002\} are detected, which arise only for the ordered regions and vanish for the disordered regions.\cite{jenichen2005,Jenichen2012}

In addition to the structural properties we also investigated the  the Co- and Ti-distributions in the films. Figure~\ref{fig:eels} (a) depicts the EELS spectrum of sample~2. The Ti-L$_{2,3}$ and the Co-L$_{2,3}$ edges are visible. Figure~\ref{fig:eels} (b) gives the plot of lateral distribution of the [Co]/[Ti]-ratio. The lateral distribution of the [Co]/[Ti]~-~ratio varies by approximately $\pm~20\%$.  This finding is confirmed by corresponding EDX measurements of the [Co]- and [Ti]-profiles (not shown here). The average [Co]/[Ti]~-~ratio measured by EDX is 2.1 (with a standard deviation of 19\% caused by the lateral variation). In principle  an angular shift of more than 1$^\circ$ could be estimated from a local lattice parameter change $\Delta$a/a~=~0.01 caused by the lateral distribution of the [Co]/[Ti]~-~ratio (cf. Figure~\ref{fig:dft_result}).\cite{jenichenEBSD}

Such lattice parameter differences are averaging already over small distances and hence do not lead to considerable broadening of the Co$_{2}$TiSi~(004) XRD peak, which is obtained on a relatively large spot of several mm$^2$. For this reason the increase of the FWHM of the measured XRD~(004) peak  is only 0.2$^\circ$ compared with the simulated curve (cf. Figure~\ref{fig:xrd004}(a)). Similar effects were observed earlier in relaxed epitaxial layer systems thanks to lateral ordering of misfit dislocations.\cite{kag97} However, the inhomogeneous distributions of the Co and Ti have an influence on the structural properties.

Figure~\ref{fig:hrtem1} (a) shows a multi-beam HR TEM micrograph of a Co$_{2}$TiSi film near an IF precipitate (sample~1). The inset depicts the fast Fourier transform (FFT) of the image and the inverted image is a magnification of the (220) maximum of the FFT. The (220) and ($\overline{2}\overline{2}$0) reflections of the FFT exhibit a peak splitting corresponding to the fully relaxed Co$_{2}$TiSi film on the GaAs substrate. As reported earlier in Ref. \onlinecite{DauAIP2015} cup-like semi-spherical precipitates (P) are found below the Co$_{2}$TiSi/GaAs interface (IF) inside the GaAs buffer layer. But more important: We find in the image of the Co$_{2}$TiSi film a structure typical for columnar growth and as a result a relatively rough surface. The diameters of the columns range from 10~nm up to 20~nm.

Figure~\ref{fig:hrtem1} (b) depicts a STEM HAADF electron micrograph of a Co$_{2}$TiSi film on GaAs exhibiting Z--contrast. The columnar structure is clearly visible starting from a film thickness of about 2.6~nm. The perfect layer near the Co$_{2}$TiSi/GaAs IF is similar to a wetting layer grown in the Stranski--Krastanov growth mode.\cite{Bauer1958} The IF and the precipitates inside the GaAs buffer layer show a brighter image than the GaAs buffer layer evidencing diffusion of the layer atoms near the IF and into the precipitates.
Figure~\ref{fig:hrtem1} (c) shows a  secondary electron micrograph of sample 1 (top view) taken in the SEM. An average column diameter of about 10~nm is visible. The perpendicular magnetization found earlier for thin Co$_{2}$TiSi films \cite{DauAIP2015} may originate from the columnar structure of the films as depicted in Fig.~\ref{fig:hrtem1}. It remains unclear up to now, why the perpendicular magnetization is observed only for thin films, maybe the columnar structure becomes too irregular in thicker films, or several magnetic domains developing over the height of the column will compensate each other.

Considering the origin of columnar growth: ordering and phase separation have been predicted for long-range interaction in ferromagnetic systems.\cite{chen1993,Dietl2008}  A "konbu" (seaweed) phase resembling a columnar structure was found for delta doping in (Zn,Cr)Te thin films.\cite{fukushima2006,Dietl2015} Here, surface diffusion during growth dominates the bulk diffusion after the growth process. In our system an increase of the [Co]/[Ti] ratio would lead to a reduced misfit (i.e. strain) between the Co$_{2}$TiSi film and the GaAs substrate. This is resulting in a driving force for composition modulation caused by epitaxial strain, i.e. the lateral separation of ordered and disordered regions accompanied by changes of the [Co]/[Ti] ratio, leading in the end to columnar growth.


\section{Summary}

Co$_{2}$TiSi films were grown by molecular beam epitaxy on GaAs(001). Evidence for columnar growth is found.  The columns open up the possibility of perpendicular magnetization of the thin film up to a thickness of 13~nm. Lateral homogeneity of the epitaxial films was characterized. We have also evidence for inter--diffusion of the different species at the interface. The films are fully relaxed and the lattice parameter of the films lies between the calculated values for \textit{L$2_1$} and \textit{B}2 ordered films.  An anisotropy of the \textit{B}2 ordering is observed as a result of the columnar growth.

\section{Acknowledgement}

The authors thank Claudia Herrmann for her support during the
MBE growth, Doreen Steffen and Sabine Krauss for sample preparation, Astrid Pfeiffer
for help in the laboratory,  Esperanza Luna for valuable support and Caroline Ch$\grave{e}$ze for critical reading of the manuscript and helpful discussion.
This work was supported in part by the Office of Naval Research through the Naval Research Laboratory's Basic Research Program. Some computations were performed at the DoD Major Shared Resource Center at AFRL. We thank the Helmholtz-Zentrum Berlin (HZB) for providing beamtime at the BESSY-beamline U125/2  KMC with the endstation PHARAO.
\newpage


\begin{thebibliography}{31}
\bibitem [{jen()}]{jenichenEBSD}%
  \BibitemOpen
  \href@noop {} {\bibinfo  {journal} {This expectation is fulfilled by the
  orientation distribution obtained by EBSD in the SEM (not shown here)
  yielding a range of orientations within 2~deg assuming a constant lattice
  parameter. However, this assumption is not justified. The result is a virtual
  tilt because in the evaluation of the measurement a homogeneous distribution
  of the lattice parameter is assumed. In this way we obtain only a seeming
  tilt, but in fact we are measuring the inhomogeneity of the lattice parameter
  due to the lateral distribution of the [Co]/[Ti]~-~ratio.}\ }\BibitemShut
  {NoStop}%
\end{thebibliography}
%

\end{document}